\documentclass[twoside,12pt]{article}
\usepackage[affil-it]{authblk}
\usepackage{epsfig}
\usepackage{graphics}

\def\r{\vec r}

\newcommand{\be}{\begin{equation}}
\newcommand{\ee}{\end{equation}}
\newcommand{\bea}{\begin{eqnarray}}
\newcommand{\eea}{\end{eqnarray}}

\begin{document}

\title{Neutrino Mass, Electron Capture and the Shake-off Contributions}

\author[1]{Amand Faessler}
\author[2]{Loredana Gastaldo}
\author[3,4]{Fedor \v{S}imkovic}

\affil[1]{Institute of Theoretical Physics, University of Tuebingen,
  Auf der Morgenstelle, D-720 76 Tuebingen, Germany}
\affil[2]{Kirchhoff Institute for Physics, Heidelberg University,
INF 227, D-691 20 Heidelberg, Germany}
\affil[3]{Department of Nuclear Physics and Biophysics, Comenius University,
  Mlynsk\'{a} dolina F1, SK-842 48 Bratislava, Slovakia}
\affil[4]{Bogoliubov Laboratory of Theoretical Physics, JINR Dubna, RU-141 980 Dubna,
Russia}

\maketitle

\begin{abstract} 
Electron capture can determine the electron neutrino mass, while the beta decay of Tritium measures the electron antineutrino mass and the neutrinoless double beta decay observes the Majorana neutrino mass. In electron capture e. g. $^{163}_{67}Ho\ + e^- \rightarrow ^{163}_{66}Dy^{*}\ +\ \nu$  one can determine the electron neutrino mass from the upper end of the decay spectrum of the excited Dy, which is given by the Q-Value minus the neutrino mass. The excitation of Dy is described by one, two \ and even three hole excitations limited by the Q-value. These states decay by X-ray and Auger electron emissions. The total decay energy is measured in a bolometer. These excitations have been studied by  Robertson and by Faessler et al.. In addition the daughter atom Dy can also be excited by moving in the capture process one (or more) electrons into the continuum. The escape of these continuum electrons is automatically included in the experimental bolometer spectrum. Recently a method developed by Intemann and Pollock was used by DeRujula and Lusignoli for a rough estimate of this shake-off process for "s" wave electrons in capture on $^{163}Ho$. The purpose of the present work is to give a more reliable description of "s" wave shake-off in electron capture on Holmium. One uses the sudden approximation to calculate the spectrum of the decay of $^{163}_{66}Dy^*$ after electron capture on $^{163}_{67}Ho$.  For that one needs very accurate atomic wave functions of Ho in its ground state and excited atomic wave functions of Dy including a description of the continuum electrons. DeRujula and Lusignoli use screened non-relativistic Coulomb wave functions for the Ho electrons 3s and 4s and calculate the Dy* states by first order perturbation theory based on Ho. In the present approach the wave functions of Ho and Dy* are determined selfconsistently with the antisymmetrized relativistic Dirac-Hartree-Fock approach. The relativistic continuum electron wave functions for the ionized Dy* are obtained in the corresponding selfconsistent Dirac-Hartree-Fock-Potential. The result of this improved approach is, that shake-off can hardly be seen in the bolometer spectrum after electron capture in $^{163}Ho$ and thus can probably not affect the determination of the electron neutrino mass.    
\end{abstract}

\section{Introduction}

The absolute values of the neutrino masses are still an open problem. Neutrino oscillations give the differences of the squared neutrino masses but not the absolute value. One hopes within the next years to obtain for the electron antineutrino mass a value or at least a better upper limit in the Tritium decay by KATRIN in Karlsruhe  \cite{Drexlin}. 

The main aim of the neutrinoless double beta decay is to distinguish, if neutrinos are of 
Dirac or Majorana nature and to measure also the effective Majorana 
neutrino mass \cite{Fae2}. 

Electron capture for example in Holmium can measure the 
electron neutrino mass \cite{de,Blaum, Alpert, Fae3}. In electron capture the upper end of the deexcitation spectrum of Dy at Q = 2.8 keV  is lowered below Q by the neutrino mass. 
The sensitivity is increased in all three methods by a small Q-value. 
\begin{eqnarray} 
 ^{163}_{67}Ho + e^- \rightarrow \  ^{163}_{66}Dy^* + \nu_e
\label{capture}
\end{eqnarray}
Energy conservation does not allow for Q = 2.8 keV to capture electrons from $^{163}Ho \ 1s_{1/2}$ with 55.618 keV, from  $2s_{1/2}$ with 9.394 keV or from  $2p_{1/2}$ with 8.918 keV binding energy (See table \ref{binding}). The first orbital,  from which an electron can be captured, is $3s_{1/2},\  M1$ with 2.128 keV binding energy.

In the sudden approximation the excitation in $Dy^*$ is given by the overlap of Holmium, with the hole due to the captured electron, and the complete set of configurations $|L,Dy>$  in Dy*. 
In case of capture from $n\ell_{1/2}$ one has:
\begin{eqnarray}
      1\  = \  <K,\  (n, \ell_{1/2})^{-1},\ Ho\ |K,\ (n, \ell_{1/2})^{-1},\ Ho> \ =  \nonumber \hspace{8cm} \\
\sum_{L,Dy} \ <K, \ (n,\ell_{1/2})^{-1},\ Ho|L,\ Dy><L,\ Dy|(n, \ell_{1/2})^{-1} , \ Ho> \ = \hspace{5cm} \nonumber \\
|<K, \ (n,\ell_{1/2})^{-1}, Ho|K,\ Dy>|^2 + \sum_{L \neq K} |<K,\ (n,\ell_{1/2})^{-1},Ho|L,\ Dy>|^2  \hspace{3cm}
\label{overlap1}
\end{eqnarray}
If one uses the Vatai approximation \cite{Vatai1, Vatai2}, setting all single electron overlaps $<nlj,Ho|nlj,Dy> = 1.0$ apart of the overlap between the captured electron orbital in Ho and the corresponding hole with the same quantum numbers in Dy, equation (\ref{overlap1}) reduces to:
\begin{eqnarray}
Prob_{shake-off} \le [1.0 - < (n, \ell,j)^{-1},\ Ho|(n,\ell,j)^{-1}, \ Dy>^{2\cdot (2j+1)}] \nonumber \\ 
\approx [1.0 -0.999^4] = 0.004 \equiv 0.4\  \%  
\label{VataiA}
\end {eqnarray}
\begin{eqnarray}
With \  10 \ \% \ error\ of \ the \   Overlap:\  < (n, \ell,j)^{-1},\ Ho|(n,\ell,j)^{-1}, \ Dy>\ \approx \ 0.9:  \nonumber \\  Prob_{shake-off} \le [1.0 - 0.90^{4}]  \approx 0.34 \equiv \ 34\ \%;   \ 34 \ \% \  of \ the \  1-hole\ states. 
 \label{VataiB}
\end{eqnarray}
Since one has only capture from $ns_{1/2}$  and $np_{1/2}$ states with j = 1/2 the exponent $2\dot(2j+1)$ is always 4. 
With electron capture in $ n, \ell_{1/2},  Ho$ the configuration K in Ho has 66 electrons as Dy, but it is not an  eigenstate of Dy. The "sum" in eqn. (\ref{overlap1}) over L in Dy represents a sum over the complete set of Dy configurations  with excitation energies less than the Q-value of 2.8 keV and  includes also an integral over the energy of the shake-off continuum states $ |E > 0, L \ Dy> $ with E positive. The second term of the last line in (\ref{overlap1}) is proportional to the probability to excite in Dy any configuration apart of the configuration K
with a hole in state $|n, \ell_{1/2}> $. This probability includes also the shake-off configurations and allows to estimate a maximal probability for the shake-off process. Thus 0.4 \% (\ref{overlap1}) and (\ref{VataiA}) is an upper limit for the shake-off probability relative to the configuration L with the 1-hole in $ n, \ell_{1/2}$. With an error of 10 \% for the overlaps and using the Vatai approximation \cite{Vatai1, Vatai2}, the shake-off can be as strong as 34 \% of the 1-hole states. Without Vatai and 10 \% error for the single electron overlaps between Ho and Dy,  the norm yields no restriction on the shake-off process. In this case shake-off  can be as large as the 1-hole states.  
\newline 
The important message from eqn. (\ref{overlap1}), (\ref{VataiA}) and (\ref{VataiB}) is, that a small uncertainty due to approximations for the electron wave functions can with this lever produce a large increase of two orders for the shake-off process. Thus very accurate wave functions are essential. 
\begin{figure}[!t]
\begin{center}
\begin{minipage}[tl]{18 cm}
\epsfig{file=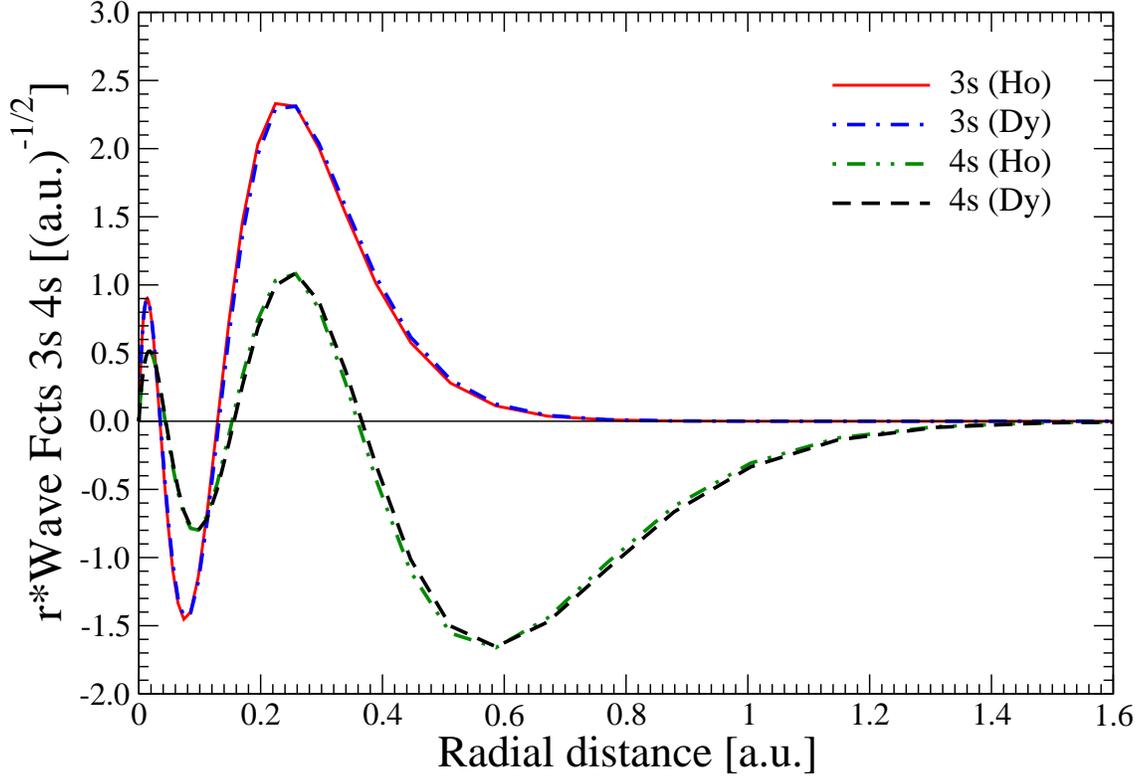,angle= -90,scale=0.6}
\end{minipage}
\begin{minipage}[t]{16.5 cm}
\caption{Non relativistic Coulomb wave functions 3s (solid), 4s (dashed-dot-dot) in Ho with the effective charges chosen by DeRujula and Lusignoli \cite{de2} $ Z_{eff}(3s)\ = \  54.9 $ and $ Z_{eff}(4s) \ = \ 43.2$. In \cite{de2} the s-wave functions for Dy including the continuum are calculated according to \cite{Int} in first order with the perturbation (\ref{pertur}).
 But for this figure the screened non-relativistic "Coulomb" wave functions for $^{163}Dy$ 
are calculated exactly \cite{Salvat} (, and not with first order perturbation relative to Ho as in ref. \cite{de2},) for the potential (\ref{pot}) with a 3s or 4s hole in Dy. The charge of the nucleus is reduced by one from Ho to Dy, but the electron hole in Dy in the orbit 3s (dashed-dot) or 4s (dashed) increases for the outer electrons the charge again effectively by one. The change from Ho to Dy is almost not visible in the wave function of this figure. The hole state has in Dy the effect to smear out the positive charge of the nucleus compared to Ho and increases by that effectively the nuclear radius. Thus the effect is to shift the functions in Dy slightly to the right to larger radial distances. This small change of the wave function restricts in the sudden approximation using the Vatai approximation \cite{Vatai1,Vatai2} the probability for 2-hole states including also shake-off to the very small values 0.00271 for 
3s and 0.00607 (see table \ref{over}) for 4s in column five of table \ref{over}. These small probabilities for 2-hole states including shake-off agrees qualitatively with the more reliable selfconsistent relativistic Dirac-Hartree-Fock \cite{Ankudinov} result of column three in table \ref{over}. This shows, that shake-off can have only a very minor effect on the bolometer spectrum of the Dy decay after capture in Ho.  
\label{HoDy3s4s}}
\end{minipage} 
\end{center}
\end{figure} 
\begin{eqnarray}
  H'(r) =\ +1/r\ -\ \int d^3r_1 \ |\varphi_{3s/4s}(\vec{r}_1)|^2/|\vec{r} - \vec{r}_1|.    
\label{pertur}
\end{eqnarray}
\begin{eqnarray}
 V(r) = -(Z_{eff}-1)/r - \int d^3r_1\ |\varphi_{3s/4s}(\vec{r}_1)|^2/|\vec{r} - \vec{r}_1|    
\label{pot}
\end{eqnarray}
The results with screened non-relativistic Coulomb wave functions including in Dy the 3s or the 4s hole state 
\cite{Salvat} and with the more exact relativistic selfconsistent electron orbitals for Dy \cite{Grant,Desclaux,Ankudinov} yield for the shake-off a smaller probability than ref. \cite{de2}, which determines the Dy orbitals by perturbation 
$H'(r)$ (\ref{pertur}) based on Ho as in \cite{Int}. Since the overlap between the corresponding Ho and Dy functions is practically 100 
 \% (see table \ref{over}) the error due to first order perturbation can only reduce the overlap. If the error of the overlaps is
  e. g. $ 10 \%$ the probability for 2-hole states including shake-off is $ 1.0 \ -\ 0.9^4\ = 0.34 $. This probability of $34\% $ is according to table \ref{over} by a about two orders larger than the correct value of about $0.4\%$. This can explain the large results for shake-off of ref. \cite{de2}. These upper limits for shake-off (\ref{overlap1}), (\ref{VataiA}) and (\ref{VataiB}) are calculated with the Vatai approximation \cite{Vatai1, Vatai2}. The norm without Vatai yields with the overlaps $\approx 0.999$ of this work an upper limit of 12\% and with a 10\% error for the in electron orbital overlaps no restriction for shake-off.
\begin{table}[!t]
\caption{Important electron binding energies in $^{163}_{67}Ho$ \cite{Williams}. Electrons can only be captured from orbitals overlapping with the nucleus. This restricts capture to $ns_{1/2}$ and in a relativistic treatment to the lower amplitude of $np_{1/2}$. Energy conservation requires, that the Q-value Q = 2.8 keV must be larger than the binding energy of the electron captured in Ho. This is the case for $3s_{1/2}$ and higher levels.}
\label{binding}
\begin{center} 
\begin{tabular}{|c |c|c|} \hline
$  n\ell_{j}$& $ Notation$ &$ E_b [keV]$\\ \hline \hline
$1s_{1/2}$&$ K1$& 55.618  \\ \hline
$2s_{1/2}$&$ L1$&  9.394  \\ \hline
$2p_{1/2}$&$ L2$&  8.918  \\ \hline
$3s_{1/2}$&$ M1$&  2.128  \\ \hline
$3p_{1/2}$&$ M2$&  1.923  \\ \hline
$4s_{1/2}$&$ N1$&  0.4324 \\ \hline 
$4p_{1/2}$&$ N2$&  0.3435\\ \hline \hline
\end{tabular}
\end{center}
\end{table}
\vspace{0.5cm}
The improvements compared to \cite{Int} and \cite{de2} are: 
\begin{itemize}   
\item  The sudden approximation \cite{Fae3,bam,Fae1, Fae4} with selfconsistent Dirac-Hartree-Fock (DHF) wave functions for the Dy atom  is used to determine the electron capture probability and not the less reliable first order perturbation theory \cite{Int,de2}.
\item  The electron wave functions in $^{163}_{67}Ho$ are not given as non-relativistic screened Coulomb functions, but are calculated with the relativistic, selfconsistent Dirac-Hartree-Fock approach \cite{Ankudinov, Grant, Desclaux}  
with full antisymmetrization. Among many other advantages the electron orbitals are in this way all orthogonal (see table \ref{over}).
\item The wave function of the bound electrons in Dysprosium are again determined selfconsistent and relativistic by Dirac-Hartree-Fock \cite{Ankudinov, Grant, Desclaux} even allowing for 3s and 4s hole states for the determination of the selfconsistent wave functions. In ref. 
\cite{Int} and \cite{de2} the electron orbitals for the daughter Dy are calculated in first order perturbation theory (\ref{pertur}). 
\item The s-wave function for the sixty-sixth continuum electron for shake-off in Dy is calculated relativistically in the selfconsistent DHF potential  of the 65 electrons in ionized Dy under the condition, that the continuum s-orbitals are orthogonal the the bound s-orbits in Dy. 
\item The problem of the numerical stability is tested carefully. For the continuum electron wave functions in Dy for the radial coordinate 250 up to more than 700 mesh points were used depending on the energy. The integration over the continuum electron energies for the shake-off electron are performed from 0 to Q = 2.8 [keV] with 417 mesh points. Integrations for the norms, the overlaps and the integration over the shake-off in the continuum were done in parallel with the Trapez rule (error $\propto$ second derivative), the Kepler-Simpson rule (error     
$\propto$ fourth derivative), the Bode-Boole rule (error $\propto$ sixth derivative) and the Weddle rule. From the points of stability and accuracy the Bode-Boole's rule  turned out to be the most reliable. All the calculations  were done in double precision. 
\item The DHF overlaps of $<3s,Ho|3s,Dy>\ = \ 0.99940$  and $<4s,Ho|4s,Dy>\  =\  0.99909$ limit in the Vatai approximation \cite{Vatai1, Vatai2} the 2-hole probability including the shake-off process, which requires a second hole, to $ 0.24 \% $ and $0.36 \%$ of the 1-hole excitations. An error of $ 10 \%$ in calculating the single orbital overlaps between  Ho and Dy due to first order perturbation theory \cite{Int,de2} estimated again with Vatai can increase the shake-off probability by two orders of magnitude. Eq. (\ref{overlap1}) serves as lever to produce from a small uncertainty  of the single electron overlaps a large increase of the shake-off probability. If one does not use the Vatai approximation and puts all electron orbital overlaps of Ho with Dy  to 0.999,  the definite upper limit (including 1- and  2-hole and shake -off excitations)  for shake-off is 12 \% relative to the 1-hole states. The norm gives without Vatai no restriction for the shake-off with an error of 10 \% for the  $<n,\ell,j,Ho|n,\ell,j,Dy>$ single electron overlaps 
\item In this work the different 1-hole, 2-hole and shake-off contributions are taken from the theory without adjusting them in different ways to fit the experiment. In ref. \cite{de2} the authors write on the second page in the right column: 
\newline " Our estimate of the height of the N1(4s)O1(5s) shakeup peak is a factor $\approx 2.5$ too low. It is possible to correct in similarly moderate ways the other contributions such as to agree with the data." 
\end{itemize}
\begin{table}
\caption{Overlaps of the 3s and 4s wave functions in Ho with Dy. The selfconsistent relativistic Dirac-Hartree-Fock results are shown in the second column. The fourth column gives the same overlaps calculated with non-relativistic screened Coulomb waves functions. The effective charges for the non-relativistic screened Coulomb wave functions are chosen as in the work of DeRujula and Lusignoli $Z_{eff} = 54.9$ for Ho 3s and excitations based on this hole state and $Z_{eff} = 43.2$ for Ho 4s and all excitations with a 4s hole. DeRujula and Lusignoli \cite{de2} use the perturbation approach of Intemann and Pollock \cite{Int} to obtain with $H' = -1/r + \int|\varphi(\vec{r}_1)|^2/|\vec{r}- \vec{r}_1|$  the wave functions in Dy. (In Dy one has one  proton less in the nucleus than in Ho and an additional electron hole in the state $\varphi(\vec{r})$.) In our  work we calculate the electron wave functions by selfconsistent DHF.  But for the numbers given in columns four and five of this table the Dy wave functions are directly calculated as non-relativistic Coulomb waves. Columns three and five give for DHF and for pure Coulomb the total probability for two hole states including also the continuum shake-off hole-particle excitations $ (ns1/2)^{-1}; (E>0, s_{1/2})^1$. These probabilities lie all well below one percent. This is smaller than found by perturbation theory \cite{de2}. The probabilities of the more exact Dirac-Hartree-Fock approach are listed in column three line two and three for 3s (M1) and 4s (N1) capture. The "non-diagonal" overlaps in lines four and five and columns two and four  are very different in the selfconsistent relativistic DHF and with Coulomb waves. In the Coulomb approach the orthogonality of different state s in the same atom is not fulfilled (lines six and seven). This is connected with the different effective charges for 3s and 4s in reference \cite{de2}. The error of the perturbation determination of the Dy wave functions can easily be 10\%. The lever of eq. (\ref{overlap1}) enlarges 10\% error in the overlap  to a 2-hole probability including shake-off $P(2-hole+shake-off)\ =\ 1.0\ -\ 0.90^4\ =\ 0.34;\ \ \rightarrow \ 34\ \%$. Comparing this values with a rough error of 10 \%  with columns three and  five it is by about two orders larger than the correct value. These upper limits increase, as discussed in the text, to 12\% and with 10\% errors to 100\%, if one does not use the Vatai approximation \cite{Vatai1,Vatai2}, but uses maximal deviations of the single electron overlaps from unity.}
\label{over}
\begin{center} 
\begin{tabular}{|c |r|r|r|r|} \hline
$  -------  $& $ DHF$ &$ 1 - <DHF>^4 $&$ Coulomb $&$ 1 - <Coul>^4$\\ \hline \hline
$<3s, Ho| 3s, Dy>$& 0.99940 & 0.00239 & 0.99932 & 0.00271 \\ \hline
$<4s, Ho| 4s, Dy>$& 0.99909 & 0.00363 & 0.99848 & 0.00607 \\ \hline
$<3s, Ho| 4s, Dy>$&-0.01982 &   ---   & 0.56828 &   ---   \\ \hline
$<4s, Ho| 3s, Dy>$& 0.02067 &   ---   & 0.56817 &   ---   \\ \hline
$<3s, Ho| 4s, Ho>$& 0.0     &   ---   & 0.56857 &   ---   \\ \hline
$<3s, Dy| 4s, Dy>$& 0.0     &   ---   & 0.56952 &   ---   \\ \hline \hline
\end{tabular}
\end{center}
\end{table}
\section{Electron Capture with Shake-off}
The deexcitation spectrum of the daughter $^{163} Dy^*$  
after electron capture in $^{163} Ho$ is described  in refs. \cite{de} and \cite{Fae3} 
assuming Lorentzian line profiles by the expression: 
\bea 
 \frac{d\Gamma}{dE_c} \propto \sum_{i = 1,...N_\nu}(Q - E_c)
\cdot U_{e,i}^2\cdot\sqrt{(Q-E_c)^2 -m_{\nu,i}^2}  \nonumber
*(\sum_{f=f'} \lambda_{0}B_f \frac{\Gamma_{f'}}{2\pi} 
\frac{1}{(E_c - E_{f'})^2 +\Gamma_{f'}^2/4} + \hspace{1cm}  \\
\sum_{f=f';p'<F;q'_b >F} \lambda_{0}B_{f,p'<F;q'_b>F} \frac{\Gamma_{f',p'}}{2\pi} \frac{1}{(E_c - E_{f',p'})^2 +\Gamma_{f',p'}^2/4}+
\hspace{3cm} \nonumber \\ 
 \int dk_{q'} \lambda_{0}B_{f,p'<F;q'_c>0} \frac{\Gamma_{f',p',q'}}{2\pi} 
\frac{1}{(E_c - E_{f',p',q'})^2 +\Gamma_{f',p',q'}^2/4} ) \hspace{4cm} \label{decay}
\eea
The factor in front of the brackets originates from the phase space of the neutrino. It is the same as for the beta decay. 
The three terms in the three lines in eq. (\ref{decay}) in the brackets describe the decay of of the excited daughter Dy from 1-hole f' excitations, from 2-hole excitations  f', p' with a shake-up of p' to q' into a bound orbit and the excitation of 2-holes f', p' with one electron p' moved to q' into the continuum with an energy $E>0$ (shake-off). The integration over the wave number $k_{q'}$ yields a dimensionless strength factor $dk_{q'} \cdot B_{f',p'<F,q'>o}$ and thus has the same dimension as the other strength factors $B_{f'} \ and \ B_{f',p'<F, q'>F}$. The transformation from an integral over the wave number to an integral over the energy  yields non-relativistically a factor $1/(2\cdot\pi\cdot k_{q'})$. (For the relativistic expression used here see eq. ( \ref{energy})).  $U_{e,i}^2$ 
is the probability for the admixture of different neutrino mass "i" eigenstates into the electron neutrino "e" flavor eigenstate. 
For the Q-value we take $Q \  = \ (2.8 \ \pm 0.08) \ [keV] $ from the ECHo collaboration \cite{Blaum, 
Anderson, Gatti, Ra, Audi}, while the recommended  
value \cite{Wang}  $Q = (2.55 \pm 0.016)$  keV seems to be to small. 
$E_c$ is the excitation energy of final Dysprosium. The energy difference $Q\ -\ E_c$ is carried away by the neutrino. $B_f $, $B_{f,p'<F;q'_b>F}$ and $B_{f,p'<F;q'_c>0}$  are the overlap and exchange corrections for the 1-hole, the bound 2-hole and the shake-off 2-hole states. $\lambda_{0}$ contains 
the nuclear matrix element squared \cite{bam}. Since $\lambda_0$ is here not calculated the theoretical results are given in arbitrary units fitted to the N1, $4s_{1/2}$ experimental peak (see figure 8).   
$ E_{f'}$, $E_{f',p'}$ and $E_{f',p';q'>0}$  are the 1-hole, the 2-hole shake-up and the 2-hole shake-off excitation energies in Dysprosium (see tables \ref{binding} and \ref{2-binding}). 
$ \Gamma_{f'}$, $\Gamma_{f',p'}$ and $\Gamma_{f',p';q'>0}$ are the widths of the one- and two-hole states and the two-hole states with shake-off  in Dysprosium \cite{Fae3,Fae4}. If the escape width of the electron in the continuum is included, it has to be added to $\Gamma_{f',p';q'>0}$ in line 3 of eq. (\ref{decay}). The escape width of the electron from the shake-off state is neglected here and in \cite{de2}. This additional escape contribution to the width should be studied in the future. It could smear out the shake-off contributions as function of the energy. The difference between the emitted neutrino and the escape electron is, that event by event the energy of the electron (plus the 2-hole binding energy) is measured in the bolometer. The neutrino escapes undetected. 
Here as in all other calculations for the deexcitation of Dy after electron 
capture a Lorentzian shape is assumed. This is probably a good description. 
Holmium is in the ECHo experiment built in a gold film positioned as an interstitial 
 or it occupies a position of the gold lattice. A Gaussian shape would be 
expected in a gas from Doppler broadening. Even collision and pressure broadening 
yield usually a Lorentzian profile. But since the shape of the resonance lines are 
important for the determination of the neutrino mass, this assumption must 
be studied in the future more carefully.  
For the neutrino mass determination the highest two hole state with an energy 2.474 keV \cite{Fae3,Robertson} is the most important excitation.  (2.0418keV 3s in Dy plus 0.4324 keV 4s from Ho. Due to the hole in Dy the second hole should "see" an effective nuclear charge similar as in Ho.)  
We describe the atomic wave function by a single Dirac-Hartree-Fock Slater determinant. 
The 1-hole $ B_f' $ and the bound 2-hole probabilities (shake-up) $B_{f',p';q'}$ are derived in refs. \cite{Fae2,Fae3,Fae4}. We concentrate here only on the shake-off probability $B_{f',p':q'> 0}$ with the electron q' in the continuum. 
The antisymmetrized Slater determinants for the  wave functions of the initial Holmium in the
 ground state $ |G> $ and the excited one electron hole states $ |A'_{f'}> $ in 
Dysprosium read in second quantization:
\be
  |G> =  a_1^{\dagger} a_2^{\dagger} a_3^{\dagger}... a_Z^{\dagger} |0> \label{G}
\ee
\be
  |A'_{f'}> =  a'^{\dagger}_1 a'^{\dagger}_2... a'^{\dagger}_{f'-1}a'^{\dagger}_{f'+1}... 
	a'^{\dagger}_{Z} |0> \label{A}
\ee
The antisymmetrized two-hole state in Dy with shake-off is:
\be
  |A'_{p',f':q'>0}> =  a'^{\dagger}_1 a'^{\dagger}_2... a'^{\dagger}_{f'-1}a'^{\dagger}_{f'+1}... 
	a'^{\dagger}_{p'-1}a'^{\dagger}_{p'+1}...a'^{\dagger}_{Z}a'^{\dagger}_{q' > 0} |0> \label{A2}
\ee
The probability to form a two-hole shake-off state is proportional to:
\be
 P_{f',p';q'>0} =  |<A'_{f', p'; q'}|a_i|G>|^2  \label{Bffff}
\ee  
The relative shake-off probability normalized to the 3s one hole excitation is: 
\begin{eqnarray}
 B_{f',p';q'>0} = \frac{|\psi_{f}(R) <A'_{f',p':q'<0}|a_f|G>|^2}{|\psi_{3s1/2}(R)|^2} 
= P_{f',p';q'>0} \cdot \frac{|\psi_{f}(R)|^2}{|\psi_{3s1/2}(R)|^2}  
\label{Bff}
\end{eqnarray}
Normally the wave function of the captured Ho electron  is taken for the nuclear matrix element at the origin. Here we take this electron wave function at the nuclear radius. Due to the weight $r^2$ of the integration this is a better choice. 
With:
\begin{eqnarray}
 P_{f',p';q'>0} = |<0|a'_{q'}a'_Z...a'_{p'+1}a'_{p'-1}...
a'_{f'+1}a'_{f'-1}...a'_{1'}\cdot a_f \cdot a^+_1...a^+_Z|0>|^2 =  \nonumber \\
|<A'_{p',f'<F; q'>F}(2\ holes)|a_f|G>|^2 
\approx   |<q'_{>0}|p_{<F}>  \cdot \prod_{k=1..Z \neq f,p}<k'|k> |^2  
\label{con}
\end{eqnarray}
q' is for the shake-off a continuum  electron orbit in Dy, into which the electron p is scattered,  
and p is the occupied state in Ho, from which this electron is removed. 
Here again k and k' and also f and f' and p and p' stand 
for the same electron quantum numbers $ n, \ \ell,\ j $ in the parent k, f, p and the 
daughter atom k', f', p'. The product over k runs 
over occupied states  $k'\ = \ k =(n_k,\ \ell_k,\ j_k,\ m_k)$ in Ho and  Dy with the exemption of 
f and p.  q' is for the shake-off contribution a continuum   
state in Dy. In the Vatai approximation \cite{Vatai1, Vatai2} one replaces the product over k in eq. (\ref{F22}) by unity. 
Because now a squared "non-diagonal" overlap is involved in eq. (\ref{con}) 
with $|<q'_{Dy}|p_{Ho}>|^2 $, the two hole shake-up and shake-off 
contributions are reduced by a "non-diagonal" overlap squared. 
If one exchanges the states f' and p', one obtains an additional "-" sign . But since one 
has to square the expression, a phase is irrelevant. 
\begin{figure}[!t]
\begin{center}
\begin{minipage}[tl]{18 cm}
\epsfig{file=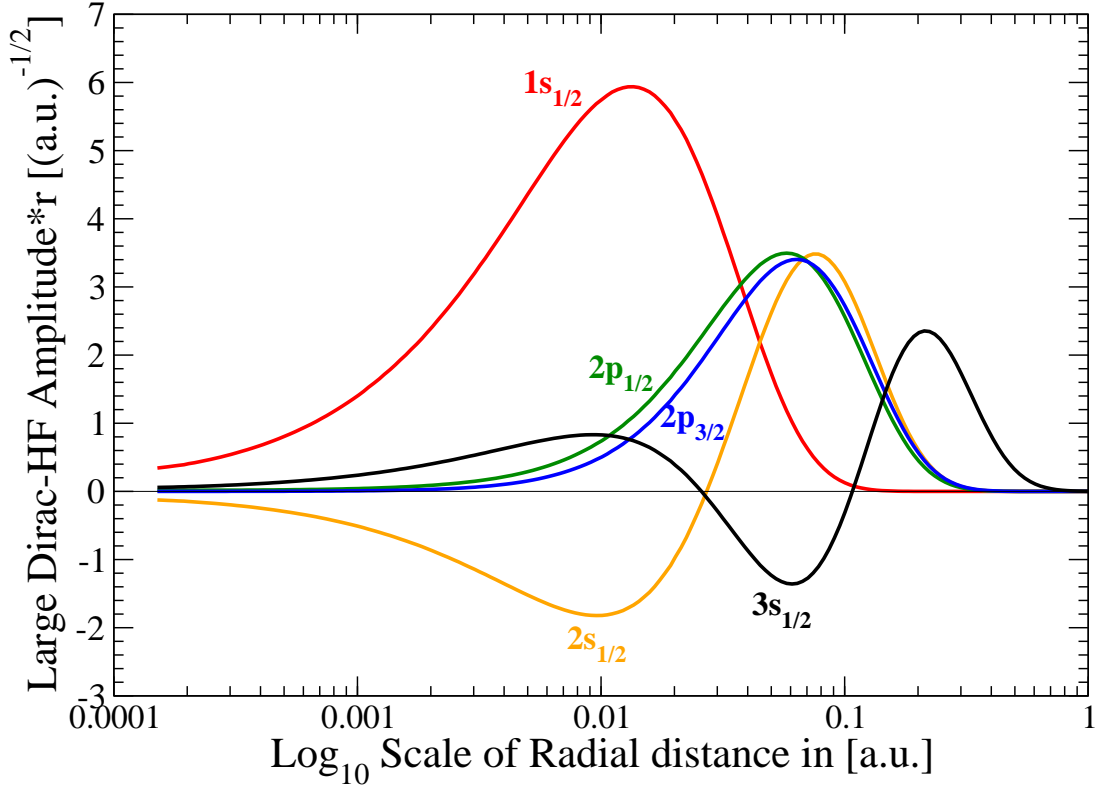,angle= -90,scale=0.6}
\end{minipage}
\begin{minipage}[t]{16.5 cm}
\caption{Large amplitudes $P(\r)$ for $1s, \ 2s, \ 2p_{1/2},\  2p_{3/2} \ and \ 3s $ Dy electrons normalized DHF wave functions in atomic units for the the radial distance (Bohr radii) and $(atomic\ units)^{-1/2}$ for the wave functions.
\label{Dy1s3s}}
\end{minipage} 
\end{center}
\end{figure} 
To evaluate the probability for the shake-off process one integrates over the wave numbers $k_{q'}$ or the the excitation energy of the continuum states q' with the same orbital  $\ell$ and total $j$ angular momentum as the state $p$ (\ref{decay}). Here the excitations are restricted to s-waves.
\bea
P_{f', p'} =  \sum_{q' > F} |<p_{<F, Ho}|q'_{>F, Dy}><q'_{>F, Dy}|p_{<F, Ho}>|
\cdot \prod_{k=k'<F_{Dy} \neq f,p}|<k'_{Dy}|k_{Ho}>|^2 
\label{F22}
\eea
In the so called Vatai approximation \cite{Vatai1,Vatai2} the overlaps 
$<k'_{Dy}|k_{Ho}> \approx 0.999$ are put to unity. 
\section{The Dy Continuum wave functions.}
To obtain the correct continuum wave functions for the shake-off electron in Dy one has three problems:
\begin{enumerate}
\item One needs a potential for this electron. This can be derived from the selfconsistent Dirac-Hartree-Fock electron wave functions in Dy taking into account the Coulomb field of the 66 protons in the nucleus and the 65 bound electrons allowing for  the different empty states. 
\item With this potential the relativistic continuum electron wave functions have to be calculated with the condition, that these states are orthogonal to all bound orbits in Dy. For calculating the relativistic wave functions in this potential we take the code of Salvat et al. \cite{Salvat} together with Schmidt orthogonalization. If the continuum waves are calculated with the non-local DHF potential, they are automatically orthogonal to the bound states like the bound orbitals among each other. Since we approximate the potential for the continuum by the local approximation (\ref{pot2}), (\ref{pot3}) in figures 3 and 4, we have to Schmidt orthogonalize the continuum wave functions (see figure 5). 
\item The wave functions of Salvat et al. \cite{Salvat} are normalized to delta functions in wave numbers $2 \ \pi \cdot \delta(k-k') $.  For the integration over the excitation energy in the continuum (\ref{decay})
 the wave functions have to be normalized to energy delta functions $\delta(E - E')$.
\end{enumerate}
\begin{figure}
\begin{center}
\begin{minipage}[tl]{18 cm}
\epsfig{file=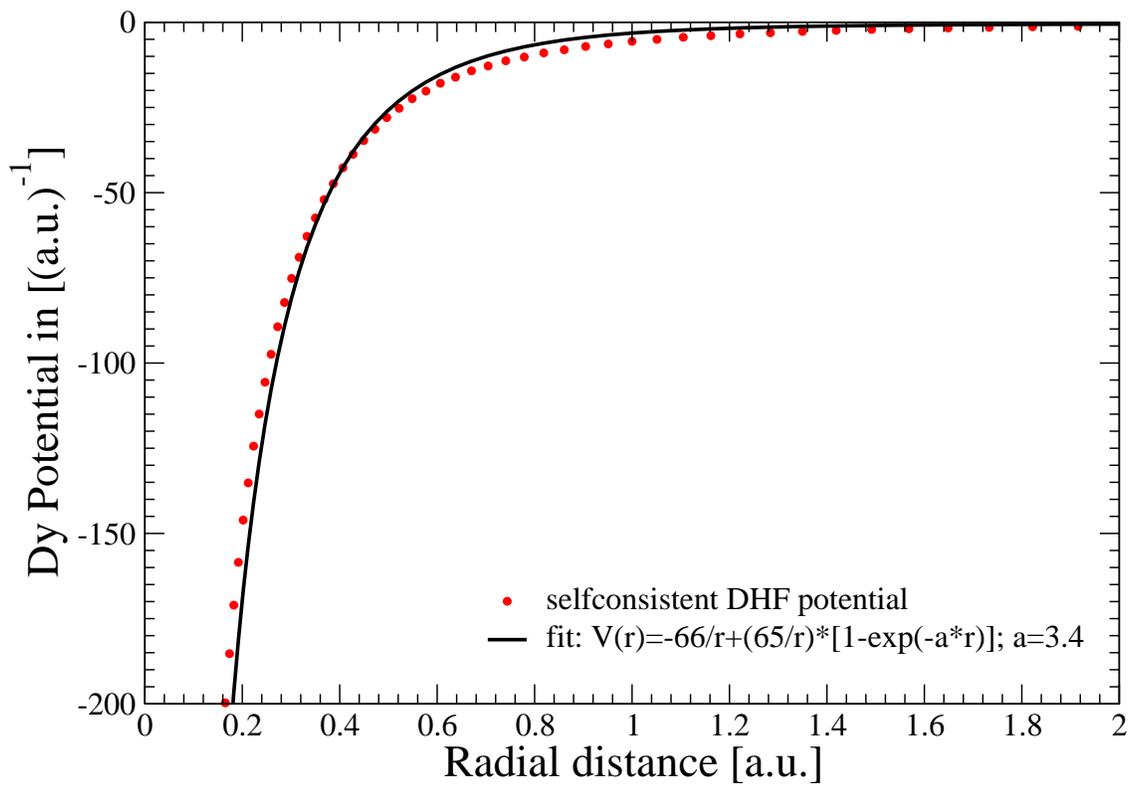,angle= -90,scale=0.6}
\end{minipage}
\begin{minipage}[t]{16.5 cm}
\caption{Selfconsistent DHF potential (dimension: 1/(length = a.u.)) and analytic approximation with a = 3.4. (See (\ref{pot2}) and 
(\ref{pot3}).)
\label{selfconsistent potential}}
\end{minipage} 
\end{center}
\end{figure}
The selfconsistent Coulomb field for the shake-off electron is in atomic units:
\be
V_{shake-off}(r) = \frac{-66}{r} +\sum_{k\ occupied \ e} g_k \int \ d^3 \r \cdot \frac{|\varphi_k(\vec{r}')|^2}{|\vec{r} - \vec{r}'|}
\label{pot2}
\ee
$g_k's$ are the number of bound electrons in the selfconsistent occupied orbits $|k>\ =\ |n,\ell,j>$  of Dy. 
To determine the potential for the shake-off electrons in Dy one needs the occupied selfconsistent Dirac-Hartree-Fock orbitals 
$P_k(r)$ and $Q_k(r)$. (As examples for P(r) see figure \ref{Dy1s3s}). 
\begin{eqnarray}
  \varphi_k(\r)*r = (\ P_k(\r) ; Q_k(\r) ) \label{PQ}
\end{eqnarray} 
\begin{eqnarray} V_{shake-off} \approx -\frac{66}{r} + \frac{65}{r} \
\cdot [1 - exp(-a\cdot r)]\ [1/(length = a.u.)] ; \ \ a = 3.4  \label{pot3} \end{eqnarray} 
We adjust an analytic expression (\ref{pot3}) to the DHF potential (see figure 3). 
At small r one obtains the Coulomb potential of the Dy nucleus $-66/r$ and for large r the dependence $-1/r$ of the ionized Dy. With the help of eqs. (\ref{pot2}) and (\ref{pot3}),  and figure \ref{selfconsistent potential} one can determine the only free parameter "a [1/(length = a.u.)]" as $a = 3.4$.
The selfconsistent Dy potential with 66 electron is shown in figure 4  for the Dy ground state and for a hole in 3s and 4s in Dy. 
\begin{figure}[!t]
\begin{center}
\begin{minipage}[tl]{18 cm}
\epsfig{file=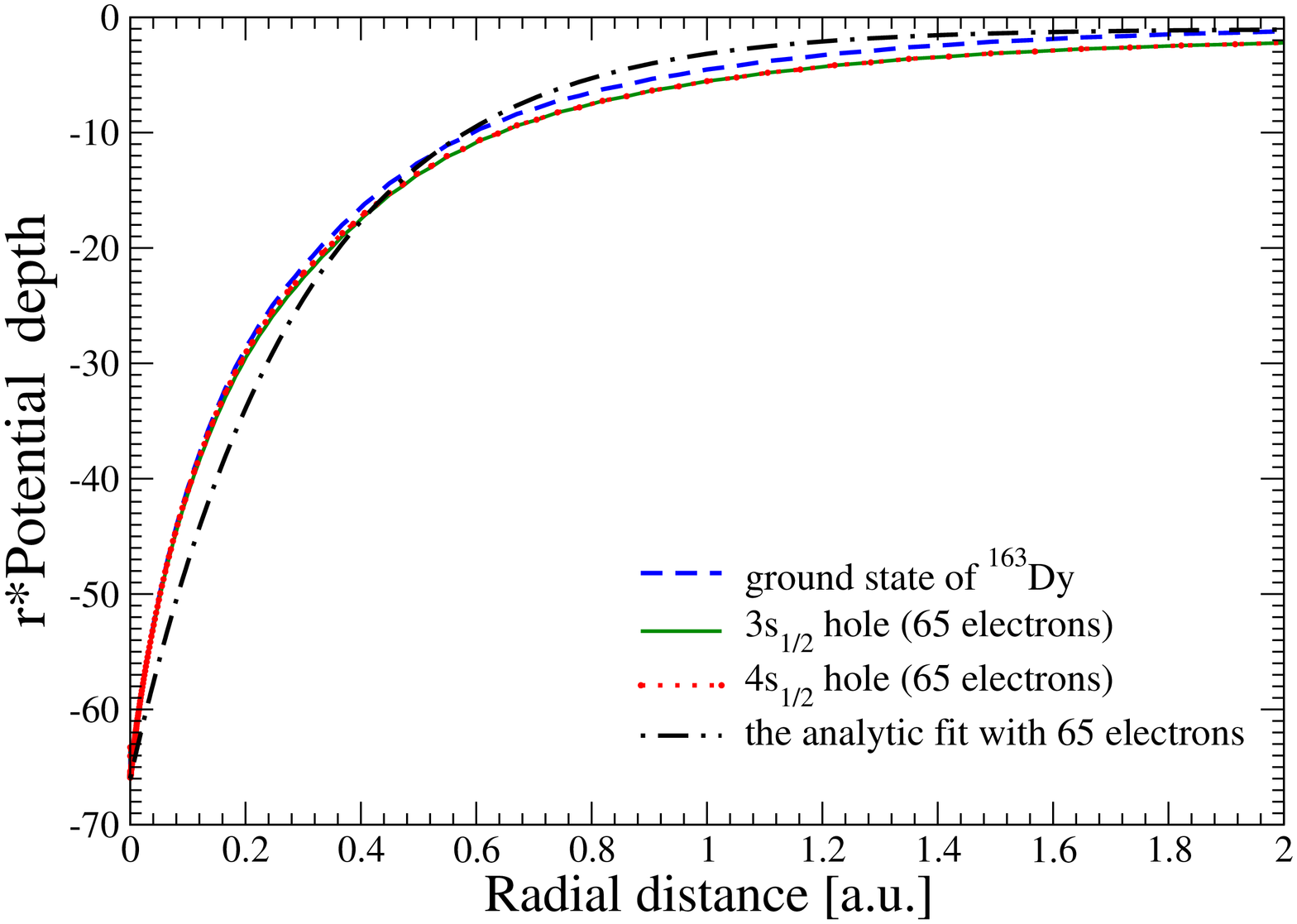,angle= -90,scale=0.6}
\end{minipage}
\begin{minipage}[t]{16.5 cm}
\caption{Radial distance r times the selfconsistent potential [dimensionless] for the ground state of the Dy atom with 66 electrons (dashed), r times the selfconsistent potential with a hole in 3s and 65 electrons (solid) , r times the selfconsistent potential with a hole in 4s and 65 electrons (dotted) and r times the analytical fit to the ground state potential (\ref{pot3}) (dashed-dot).
\label{Dy-potential}}
\end{minipage} 
\end{center}
\end{figure} 
\begin{figure}[!t]
\begin{center}
\begin{minipage}[tl]{18 cm}
\epsfig{file=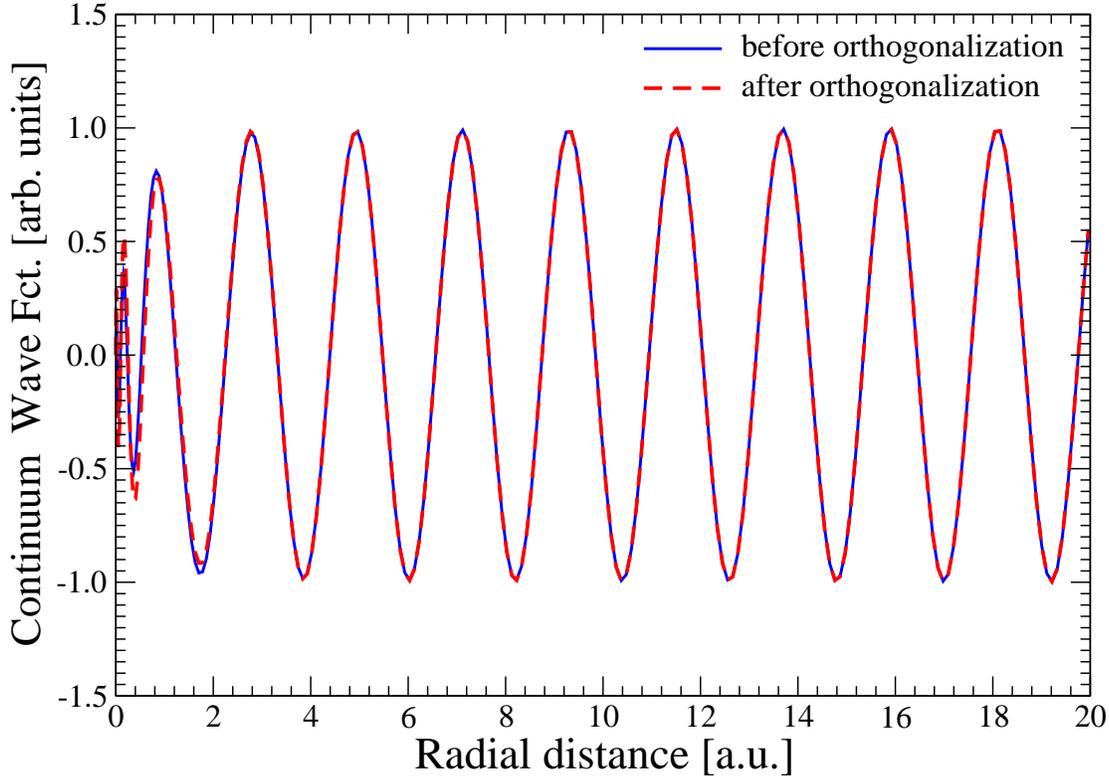,angle= -90,scale=0.6}
\end{minipage}
\begin{minipage}[t]{16.5 cm}
\caption{Large amplitude $P(r)$ of the s wave at $E \ = \ 4 \ [Hartree] 
\ \equiv 108.8 \ [eV]$ before and after Schmidt orthogonalization.
\label{continuum}}
\end{minipage} 
\end{center}
\end{figure}
\begin{figure}[!t]
\begin{center}
\begin{minipage}[tl]{18 cm}
\epsfig{file=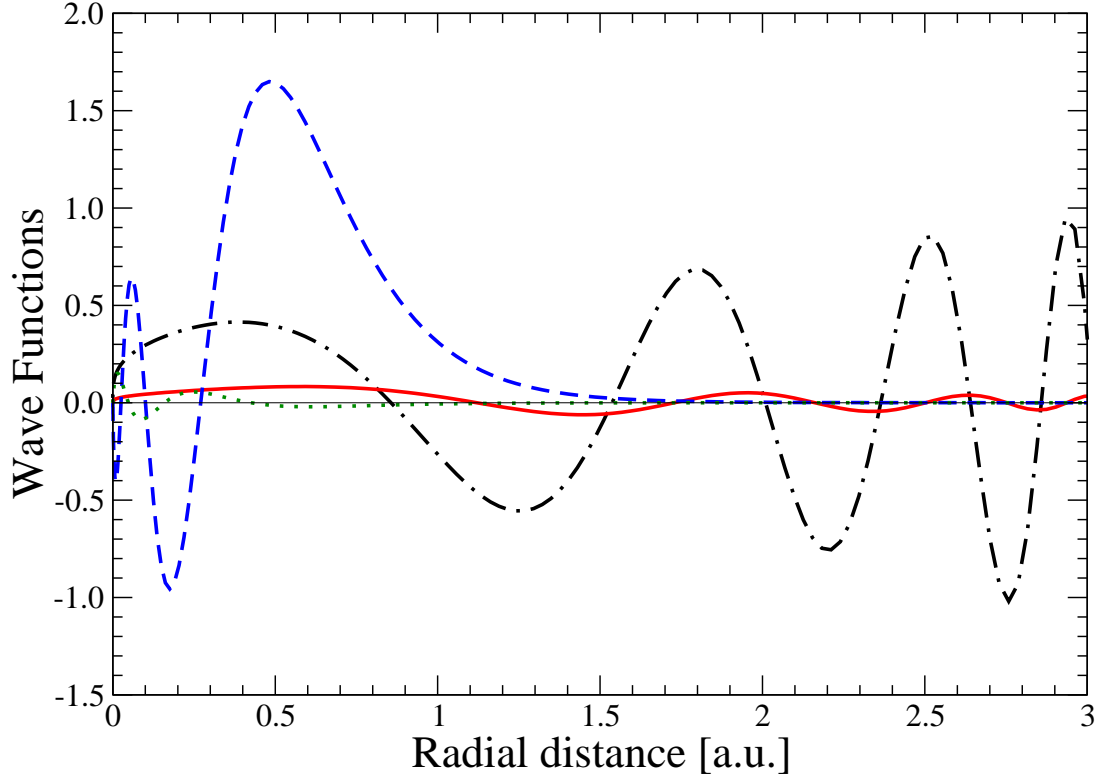,angle= -90,scale=0.6}
\end{minipage}
\begin{minipage}[t]{16.5 cm}
\caption{Wave function P(r) (dashed-dot) and Q(r) (solid) (\ref{PQ})  in the Dy continuum at 50 [Hartree] = 1.36 keV. The continuum wave functions P and Q are normalized to $\int  dr [P(r)^2 + Q(r)^2\ ] = \  2\pi \ \cdot  \delta (k - k')$. The 4s bound state in Ho is normalized to unity (dashed for P(r) and dotted for Q(r)): $\int dr [P(r)^2 + Q(r)^2 ] = 1 $. The overlap $<ns,\ Ho|E, \ s,\ Dy> $ squared is proportional to the shake-off process as a function of energy. The continuum wave functions $P_E\  and\  Q_E$ are dimensionless, while the bound states $ P_{4s}(r)\  and\  Q_{4s}(r)$ are in atomic units $[(a.u.)^{-1/2}]$.
\label{Dycontinuum-Ho4s}}
\end{minipage} 
\end{center}
\end{figure}
\begin{figure}[!t]
\begin{center}
\begin{minipage}[tl]{18 cm}
\epsfig{file=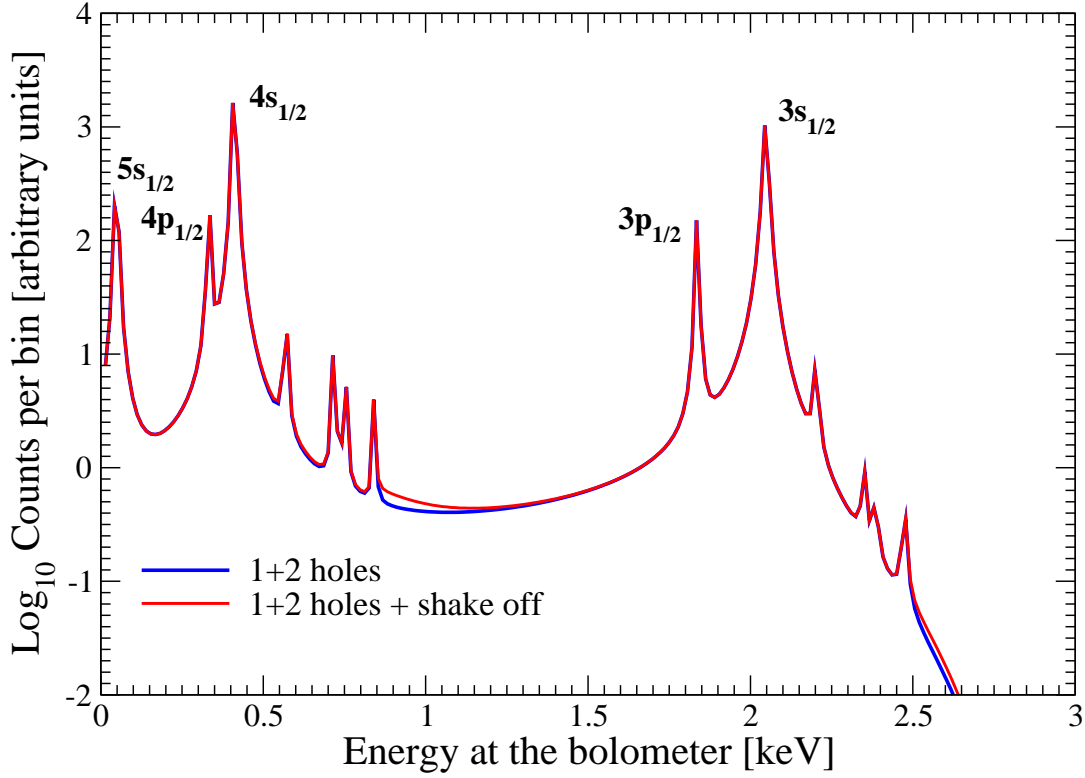,angle= -90,scale=0.6}
\end{minipage}
\begin{minipage}[t]{16.5 cm}
\caption{Theoretical results in arbitrary units of the sum of the one- and two-hole deexcitations compared to the sum of the one-, two-hole and the shake-off deexcitation as measured by the bolometer spectrum (\ref{decay}). The arbitrary units are adjusted to the experimental N1, $ 4s_{1/2} $ 1-hole peak (see figure 10). The nature of the one hole states are indicated. The two-hole peaks are by about two orders of magnitudes smaller than the one hole peaks. Shake-off can almost not been seen.  
\label{spectrum}}
\end{minipage} 
\end{center}
\end{figure}
\begin{figure}[!t]
\begin{center}
\begin{minipage}[tl]{18 cm}
\epsfig{file=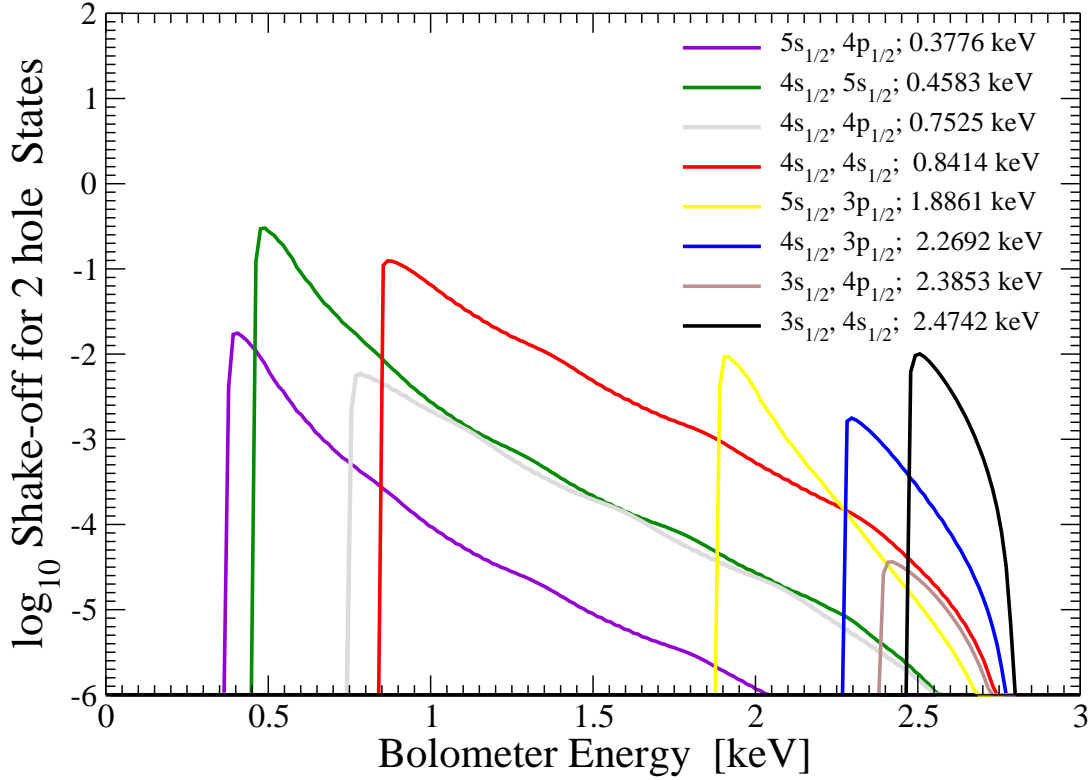,angle= -90,scale=0.6}
\end{minipage}
\begin{minipage}[t]{16.5 cm}
\caption{Shake-off contributions for different 2-hole excitations in Dy normalized for the experimental bolometer spectrum (see figure 10)  to the N1, $4s_{1/2}$ peak.  Increasing the energy $E_c$ of the bolometer spectrum  the Q-value = 2.8 keV is first used to excite the two hole state. So the shake-off contribution for the bolometer spectrum starts as function of $E_c$ with the 2-hole binding energy. Energy conservation yields an upper limit of Q = 2.8 keV for the bolometer spectrum. To integrate over the continuum energy of the shake-off electron (\ref{decay}) 
we divided the interval $<0.0\ ; \  2.8\ keV> $ into 417 mesh points.  From the left to the right with increasing bolometer 
energy $E_c$  the start of the different shake-off contributions are indicated in the figure. The energy difference between Q and $E_c$ is carried away by the neutrino, which can not contribute to the bolometer spectrum. 
\label{shake-off}}
\end{minipage} 
\end{center}
\end{figure}
\begin{figure}[!t]
\begin{center}
\begin{minipage}[tl]{18 cm}
\epsfig{file=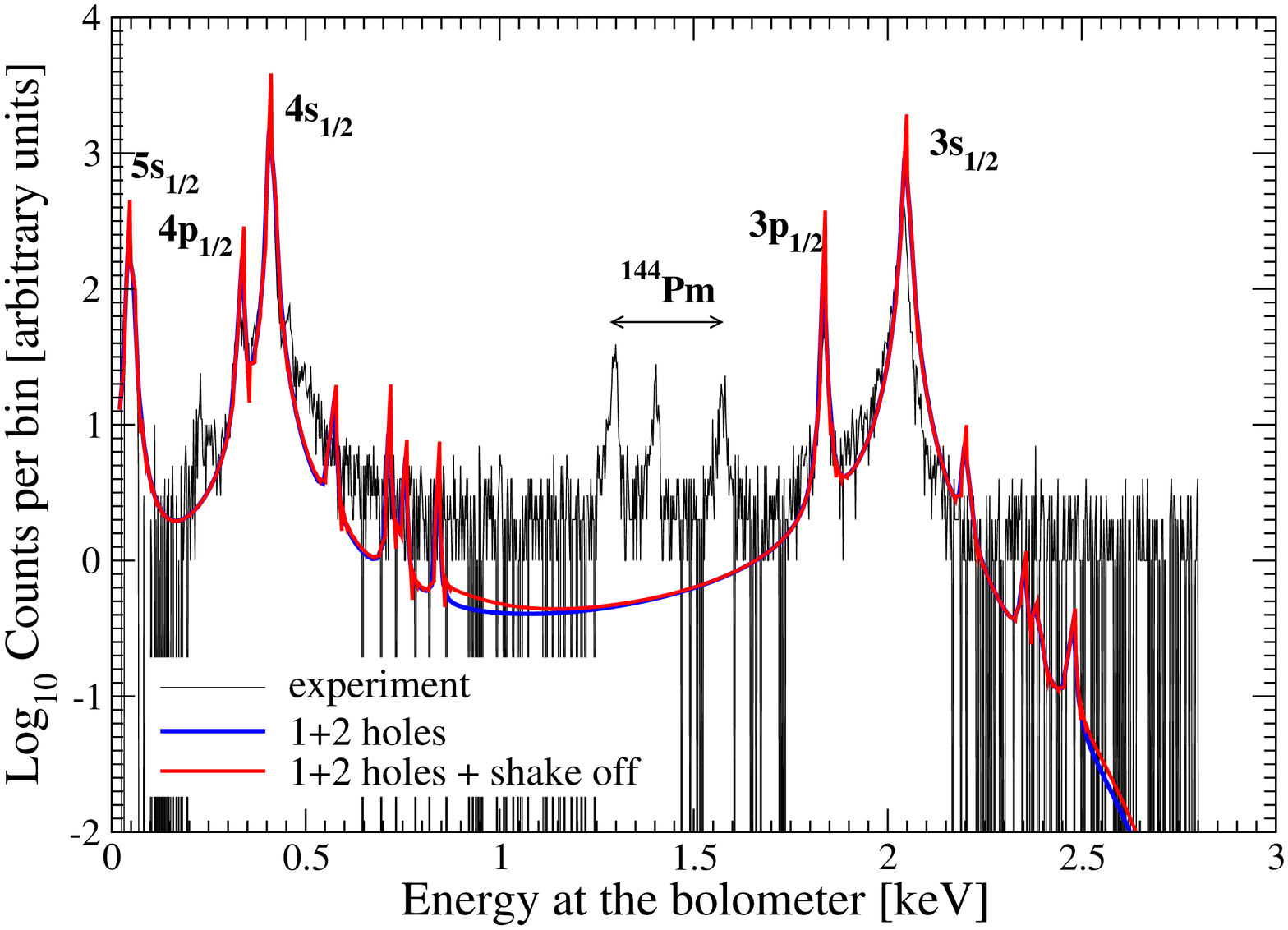,angle= -90,scale=0.6}
\end{minipage}
\begin{minipage}[t]{16.5 cm}
\caption{ Experimental and theoretical results of the sum of the one- and two-hole deexcitations and the sum of the one-, two-hole and the shake-off deexcitation for the bolometer spectrum (\ref{decay}). The experimental data are from the ECHo collaboration 
\cite{Blaum} and \cite{Ra}. The two theoretical spectra are adjusted to  experiment at the  $N1, \ 4s_{1/2}$ peak. The nature of
the one hole states are indicated. The two-hole peaks are by about two orders of magnitudes smaller than the one hole peaks. The shake off contributions can hardly been seen in this scale. Some bins contain no experimental counts, thus the $log_{10}$ for these experimental values are minus infinity. To fit the 1931 experimental points for the bolometer energy of 0.0 to 2.8 keV, the theoretical spectrum of 200 mesh points had to be interpolated to the data points for this figure. Figure 7 contains the 200 original theoretical results without interpolation for the bolometer spectrum over $E_c$  between 0.0 and 2.8 keV. The interpolation is normally very good (compare figures 7 and 9) but difficult at some sharp minima and maxima. 
\label{spectrum10}}
\end{minipage} 
\end{center}
\end{figure}
The Relativistic continuum wave functions in the potential (\ref{pot3}) are determined with the code of Salvat, Fernandes-Varea and Williamson \cite{Salvat}. This code can handle potentials with the properties $ \lim_{r \rightarrow \infty} r \cdot V(r) = constant$. All our potentials are of this nature. The wave functions are normalized with the WKB approximation to:
\begin{equation} \lim_{r \rightarrow \infty}\  P_k(r)\  = \ 2\cdot\sin(kr - \ell \cdot \frac{\pi}{2} -\eta \ln 2kr+ \Delta +\delta) 
\label{asymptotic} \end{equation} 
For Q(r) one has a similar asymptotic expression. The problem of the normalization of the continuum Dirac wave function is for example discussed by M. E. Rose in his book on "Relativistic electron theory " \cite{Rose} or by Walter Greiner in the book on "Relativistic Quantum Mechanics" \cite{Greiner}. The normalization is also discussed by Goldberg et al. \cite {Goldberg}, and by Perger and Karighattam \cite{Perger} on their page 394. We follow here this recommendation \cite{Perger}. In the asymptotic expression  (\ref{asymptotic}) $\eta$ is the Sommerfeld parameter, $\Delta$  the Coulomb phase shift and $\delta$ takes into account deviations from a pure Coulomb potential. The usual way \cite{Perger} to determine the norm is to normalize only P(r) to the delta-function in wave numbers or energies and treat the small relativistic amplitude in the same way. Since the electron energies required for shake-off in the continuum are small compared to the electron rest mass $m_e c^2\  = \ 510.9989 keV$ this often used normalization should be good for our purpose. The electron energy in the continuum can be due to energy conservation not larger than the Q-value of 2.8 keV minus the excitation energy of the two hole state. So for shake-off with capture from 3s the energy of the electron in the continuum must be less than 0.758 keV (extreme non-relativistic). The most  important second hole is $4s1/2$ and thus for the two holes in 3s and 4s the integration over the shake-off electrons is restricted to $2.800 -2.474 keV = 0.236 keV$. For capture from the two $4s1/2$ states the binding energy limits the integration in eq. (\ref{decay}) to an upperlimit of $2.800\  -\ 0.841\  = \ 1.959\  keV$. 
\\ \\
The asymptotic (\ref{asymptotic}) requirement normalizes the asymptotic form of P(r) with $\lim_{r \rightarrow \infty}\  P_k(r)\ $ to the delta-function in wave numbers. Continuum wave functions for different electron energies are orthogonal. The delta function strength is determined by the asymptotic, which yields infinity for the norm integration. Thus for continuum wave functions (\ref{asymptotic}) calculated with the DHF selfconsistent local potential (\ref{pot2}) and (\ref{pot3}) is also normalized in the wave numbers to $2\cdot \pi \cdot \delta(k - k')$ and in the energy to $\delta(E -E')$. The Schmidt orthogonalization does not change this, since it modifies the wave functions only at short distances and the delta function is determined by the asymptotic. Comparing the two asymptotic forms gives us the transformation factor from the wave number $2\cdot \pi \cdot \delta(k - k')$ to the energy delta function $ \delta(E - E')$ normalization.
\newpage
\begin{eqnarray} \int_{r = 0 \ to \ \infty}  dr \cdot 2\cdot \sin(kr - \ell \cdot \frac{\pi}{2} -\eta \ln 2kr+ \Delta +\delta) \cdot \\ 2\cdot\sin(k'r - \ell \cdot \frac{\pi}{2} -\eta \ln 2kr+ \Delta +\delta) \approx 2\pi \cdot \delta(k-k'). \nonumber 
\label{knorm} \end{eqnarray} 
\\
The wave number is connected with the relativistic and non-relativistic energies by the equations:
\\
\begin{eqnarray} E^2_{rel} = c^2 \hbar^2 k^2 + m^2 c^4 \ \rightarrow \ c^2 k^2 + c^4\ \ (in\  atomic \ units);\ \  E_{n-rel} = \frac{1}{2}k^2; \hspace{1.5cm} \nonumber  \\ 
 k = \alpha \sqrt{E_{n-rel}(E_{n-rel} +  2\cdot c^2)}; \ \ \ with \ c \ = \ 1/\alpha \ = \ 137.035999\  in\ [atomic\  units].
\label{wavenumber} \end{eqnarray} 
\\
We use to change from the asymptotic wave number normalization $2\pi \cdot \delta(k-k')$  to the normalization $\delta(E - E')$  well known relations for Dirac delta functions given for example in Landau-Lifschitz, 
 Volume 3, "Quantum Mechanics" chapter 5 and chapter 33 \cite{Landau}. 
\begin{eqnarray} \delta(g(x)) = \frac{\delta(x - x_0)}{|g'(x_0)|}  \nonumber \\
\delta(a\cdot x) = \frac{1}{|a|}\cdot \delta(x) \\
with: g(x_0) = 0
\label{delta}    \nonumber
\end{eqnarray} 
\\
The transformation from the asymptotic wave number normalization $P_k(r)$ to the asymptotic energy delta function normalization is:
\begin{eqnarray} P_E(r) = P_k(r) \cdot \sqrt{\frac{1}{2\pi} \cdot  \frac{1}{c} \cdot \frac{ \sqrt{k^2+c^2}}{k}}\ \approx P_k(r)\cdot \frac{1}{\sqrt{2 \cdot \pi\ \cdot k}} \hspace{1cm} \\ \nonumber
\delta(E[Hartree])\  =\  \delta(36.74932386 \cdot E[keV])\  = \ 0.027121138506 \cdot \delta(E[keV])
\label{energy} 
\end{eqnarray} 
\\
We transform the $Q_k(r)$ in the same way using the asymptotic expression to obtain the transformation factor. 
For shake-off one has to calculate the overlap of the bound Ho electron orbitals $|p>$ with the continuum wave functions in Dy $|q'>$ i.e.
$<q'_{>0}|p_{<F}>$ (\ref{con}) (see figure 6). For this one expands the configuration of Ho after capture of the bound electron $|b,Ho>$ with now the same number of protons as Dy (- but not a Dy eigenstate -) into the complete set of configurations $|L, Dy>$  in Dy including the continuum.
\newpage
\begin{eqnarray} 
|K, \ (b)^{-1}, \ Ho> = \sum_{L \neq K \ , bound} a_{L} \cdot |L,\ Dy> + \int_{0\ to\ \infty} dE' \cdot a(L,\ E') |L, \ E',Dy> \hspace{2 cm} \nonumber \\
a_{L} = <L ,Dy|(b)^{-1},Ho>, \hspace{8cm} \nonumber \\
a(L, \ E'') = <L, E'',Dy|(b)^{-1}, Ho> =  \hspace{6cm} \nonumber \\
\int_{0 \ to\ \infty} \cdot dE' \cdot a(L, E') \cdot <L, E'',Dy|L, E',Dy> \hspace{4cm}
\label{expansion}
\end{eqnarray} 
Here the delta-function normalization of the continuum wave functions in Dy is used: $<E'',Dy|E',Dy> = \delta(E'' -E')$. 
The probability forming a specific hole state $|k',Dy>$ in Dy in a bound orbit after capture of the electron $|b,Ho>$ is proportional to $|<k',Dy|b,Ho>|^2$ and and for the continuum $|E'',Dy>$ to $|<E'',Dy| b, Ho>|^2$ 
integrated over the continuum energy  of the shake-off electron E''. 
\section{Results for Shake-off.}
To calculate the shake-off contributions, one has to determine first the overlap between the bound Ho states $ns_{1/2}$ and in principle also $np_{1/2}$ with $n \ge 3$ and the continuum wave functions in Dy. Since we restrict this work to s-wave shake-off we need only the overlaps $<n \ge3, s_{1/2}, Ho|E,\ s, Dy>$. In the summed spectrum with 1-hole, 2-hole and shake-off in figure 7 and 9  the shake-off contribution is hardly  visible. Here and also in ref. \cite{de2} only the decay width of the 2-hole states are included. Three hole states can be neglected \cite{Enss}. The electron in the continuum has an  escape width, which is not included. 
\begin{table}[!t]
\caption{Electron binding energies and width of two-hole states in $^{163}_{66}Dy$, which contribute to s-wave shake-off. Energy conservation requires, that the Q-value Q = 2.8 keV must be larger than the two-hole binding energy plus the energy of the electron in the continuum. The shake-off contributions for the 2-hole states start at the2-hole binding energy in the bolometer spectrum as a function of $E_c$. The width includes only the contribution from the decay of the 2-hole states, but not the escape width of the continuum electron.}
\label{2-binding}
\begin{center} 
\begin{tabular}{|c |c|c|} \hline
$ n_1\ell_{j,1}, n_2\ell_{j,2} \ $& $ 2-hole\  E_b [keV]$&$ Width\ [keV]\ $\\ \hline \hline
$3s_{1/2}, 4s_{1/2}$& 2.4742& 0.0264 \\ \hline
$4s_{1/2}, 4s_{1/2}$& 0.8414& 0.0108 \\ \hline
$4s_{1/2}, 5s_{1/2}$& 0.4583& 0.0107 \\ \hline
$4s_{1/2}, 3p_{1/2}$& 2.2692& 0.0114 \\ \hline
$5s_{1/2}, 3p_{1/2}$& 1.8861& 0.0114 \\ \hline
$3s_{1/2}, 4p_{1/2}$& 2.3853& 0.0186 \\ \hline 
$4s_{1/2}, 4p_{1/2}$& 0.7525& 0.0107 \\ \hline
$5s_{1/2}, 4p_{1/2}$& 0.3776& 0.0106 \\ \hline \hline
\end{tabular}
\end{center}
\end{table}
\begin{table}
\caption{Overlaps of $^{163}Ho$ electron orbits with bound and continuum wave functions (\ref{knorm}) $P_k(r)$  and $Q_k(r)$ in $^{163}Dy$. The continuum wave functions are normalized asymptotically as in eq. (\ref{asymptotic}) to the delta function for the wave numbers (\ref{knorm}). For the integral over the continuum energy one has to square the overlaps of Ho functions with the continuum in Dy and to change to the energy normalisation ( see after eqn. (\ref{energy})). This transformation squared gives roughly a factor: $1/(2\pi\cdot k) \approx 0.014\ [a.u.]$ for an electron energy of 1.768 keV in the Dy continuum. On the other side the transformation from the delta function of energies in Hartree to energies in keV increases shake-off result by a factor 36.7498 [Hartree/keV].}
\label{Overlaps} 
\begin{center}
\begin{tabular}{|c |c|c|c|} \hline
$  --- $ &$ |3s, Dy>$&$ |4s,Dy>$&$ |5s,Dy> $\\ \hline \hline
$ <3s, Ho|$&$ 0.9940$&$  -1.98226\cdot10^{-2} $&$  6.20190\cdot10^{-3}$      \\ \hline]
$ <4s, Ho|$&$ 2.06722\cdot10^{-2}$&$  0.99909 $&$ -1.87503\cdot10^{-2}$      \\ \hline
$ <5s, Ho|$&$-6.31396\cdot10^{-3}$&$  1.97766\cdot10^{-2}$&$ 0.99928$     \\ \hline \hline 
$ --- $&$|E=0.884 keV,s,Dy>$&$ |E=1.768 keV,s,Dy>$&$ |E=2.653 keV,s,Dy> $      \\ \hline \hline
$ <3s, Ho|$&$ 1.383\cdot10^{-2}$&$  1.044\cdot10^{-2}$&$ 8.314\cdot10^{-3}$  \\ \hline
$ <4s, Ho|$&$-1.407\cdot10^{-2}$&$  -7.366\cdot10^{-3}$&$ -4.881\cdot10^{-3}$\\ \hline
$ <5s, Ho|$&$ 4.926\cdot10^{-3}$&$  2.095\cdot10^{-3}$&$ 1.329\cdot10^{-3}$  \\ \hline \hline 
\end{tabular}
\end{center}
\end{table}
Figure 7 shows the logarithmic spectrum of the 1-hole, the 2-hole and the s-wave shake-off contributions. The shake-off contributions are calculated for the different 2-hole states listed in table \ref{2-binding}. The two-hole spectrum is about two orders of magnitude smaller than the one-hole states. The shake-off spectrum can hardly  be seen on this scale in the total spectrum. Compared to the one-hole peaks it is at least two orders smaller. The integration over the continuum electron energy (\ref{decay}) is  done by the Bode method using 417 mesh points.  Shake-off is proportional to the square of the overlap $<Ho-bound|Dy-continuum> $. The 2-hole states contributing to s-wave shake-off are listed in table \ref{2-binding}.  The shake-off  contributions of the 2-hole states as function of the bolometer energy $E_c$ is starting from the 2-hole binding energy up (see table \ref{2-binding}). The two main contributions originate from 4s1/2, 5s1/2 starting at 0.4583 keV and from 4s1/2, 4s1/2 starting at 0.8414 keV. The $log_{10}$ contributions from 3s1/2, 4p1/2 starting at 2.33853 keV (see table \ref{2-binding}) are extremely small. The energy difference $Q - E_c$ is carried away by the neutrino and and does not show in the bolometer. 
\section{Conclusions}
In this work the effect of shake off on the deexcitation  spectrum of the $^{163}_{66}Dy^*$ atom after electron capture in 
$^{163}_{67}Ho$ for the determination of the electron neutrino mass is investigated. The electron neutrino mass is the difference between the upper end of the deexcitation spectrum of Dy* measured by a bolometer and the Q-value. After capture the Dy* can be excited into 1-hole and into 2-hole electron configurations. The three hole excitations and higher  can be neglected \cite{Enss}. The total 2-hole excitation probability 
is given in the sudden plus the Vatai \cite{Vatai1, Vatai2} approximation by unity minus the overlap squared between Ho and Dy with the number of electrons in the exponent with the same quantum numbers in Dy as in Ho for the orbit, from which the particle is captured (\ref{overlap1}) and (\ref{VataiA}).
\be  (1.0 - <Ho, n, \ell,j|Dy,n,\ell,j>^{2(2j+1)})  \label{overlap2} \ee
These Ho-Dy overlaps have in selfconsistent relativistic Dirac-Hartree-Fock values of about 0.999 and even closer to unity  (see table \ref{over}). Thus  this total 2-hole probability including shake-off must be according to this rough estimate  less the $0.4 \ \%$. This estimate is only very approximate. Important is the fact, that a small uncertainty of for example 10 \% for this overlap produces a large increase of about two orders of magnitude for shake-off. Without using Vatai \cite{Vatai1,Vatai2} the upper limit for shake-off increases in our work to 12\% and for a 10\% error in the overlaps to 100\%.  The excited Dy* wave functions are calculated in a previous investigation \cite{Int, de2} in first order with the perturbation (\ref{pertur}) from the Ho states. An uncertainty in the overlap $<ns_{1/2}, Ho|ns_{1/2},Dy> $ between Ho  and Dy for the sudden approximation reduces always the overlap.   A $10 \ \%$ error in the overlap calculated with perturbed Dy*  wave functions based on pure Coulomb waves in Ho can produce an overestimation of shake-off by about  two orders of magnitude. The bound states in Ho and Dy are described here by the  Dirac-Hartree-Fock approach \cite{Ankudinov, Grant, Desclaux} even including different occupations in Dy due to the hole states. The s-wave continuum wave functions in Dy are determined with the Dirac equation in the selfconsistent potential \cite{Salvat}. The energy of the continuum states involved are limited by energy conservation to the Q-value minus the 2-hole binding  energies. E. g. the $ 3s_{1/2},\  4s_{1/2}$ 2-hole state limits the upper bound of the continuum energy contributions to 2.8 - 2.4742 = 0.3258 keV. So this contribution is very small. One of the two  main contributions comes from $4s_{1/2},\  4s_{1/2}$ with a binding energy of 0.841 keV and thus an upper limit of the continuum energy of 2.8 -  0.841 = 1.959 keV. A second large contribution is built on the 2-hole state $ 4s_{1/2},\  5s_{1/2}$ with the binding energy 0.4398 keV. Thus the upper limit of the shake-off contributions in the continuum integration (\ref{decay}) is the 2.8 - 0.4398 = 2.3602 for the integral over the shake-off continuum electron. 
\newline 
We prepared two different computer programs  both calculating the s-wave shake-off to test the two codes against each other. All the calculations are done in double precision and for the critical integrations we use parallel the Trapez, the Kepler-Simpson, the Bode and the Weddle rules to test the accuracy. The numbers given are the ones from the Bode rule. (Trapez is not reliable enough.) The contributions  from the shake-off process are small (see figures 7 to 9). The widths for the shake-off states include only the values from the 2-hole excitations as in \cite{de2}. In reality one has to include the escape width of the electron in the continuum, which could perhaps even be larger than the contribution of the 2-hole states.
\newline
In summary this work shows, that one has not to worry about the shake-off process in the determination of the neutrino mass from electron capture in  $^{163}Ho$. 
\newline 
The remaining discrepancies between theory and experiment, e. g. the slope above the 1-hole state $4s_{1/2} \ (N1)$, are probably due to configuration mixing not included  here. 
Finally we want to stress, that the accuracy needed to extract the neutrino mass  can not be obtained by theoretical calculation alone. One must fit the neutrino mass, the Q-value, the highest resonance hole energy and their width at the upper end of the spectrum to extremely accurate data.  
\vspace{0.5cm}
F. \v{S} acknowledges the support in part by the Heisenberg-Landau program the VEGA Grant Agency
of the Slovak Republic under Contract No. 1/0922/16.
\vspace{1cm}
\end{document}